\documentclass[pre,showpacs,byrevtex,amsmath,amssymb,preprint]{revtex4}

\usepackage{graphicx}
\usepackage{dcolumn}
\usepackage{bm}

\begin{document}
\preprint{PREPRINT}

\title[Short Title]{Effect of Image Forces on Polyelectrolyte Adsorption at a Charged Surface
                  }

\author{Ren\'e Messina}
\email{messina@thphy.uni-duesseldorf.de}
\affiliation
{Institut f\"ur Theoretische Physik II,
Heinrich-Heine-Universit\"at D\"usseldorf,
Universit\"atsstrasse 1,
D-40225 D\"usseldorf,
Germany}

\date{\today}

\begin{abstract}
The adsorption of flexible and highly charged polyelectrolytes onto oppositely charged planar 
surfaces is investigated by means of Monte Carlo simulations. 
The effect of image forces stemming from the 
dielectric discontinuity at the substrate interface is considered. 
The influence, at fixed polyelectrolyte volume fraction, of chain length and surface-charge
density is also considered. A detailed structural study, including monomer and fluid 
charge distributions, is provided. 
It is demonstrated that image forces can considerably reduce 
the degree of polyelectrolyte adsorption and concomitantly
inhibit the charge inversion of the substrate by polyelectrolytes.
\end{abstract}
\pacs{82.35.Gh, 82.35.Rs, 61.20.Qg, 61.20.Ja}
\maketitle

\section{Introduction}

The adsorption of charged polymers [polyelectrolytes (PEs)]
on charged surfaces is an important phenomenon
in industrial and biological processes.
Recent model and well controlled experiments 
\cite{Hansupalak_Langmuir_2003,Tulpar_JPCB_2004}
were devoted to characterize PE adsorptions.
The understanding of PE adsorption remains an outstanding
problem because of the many different typical interactions involved there:
strong electrostatic substrate-PE binding, monomer-monomer (PE-PE) repulsion, 
chain-entropy, excluded volume, etc. 
Another complication arises from the \textit{dielectric discontinuity} between the
solvent and the substrate generating surface-polarization charges, which
occurs in experimental situations.  

On the theoretical side, PE adsorption on planar charged surfaces has been 
intensively studied by several authors  
\cite{Wiegel_JPhysAMathGen_1977,
VanderSchee_JPC_1984,
Muthukumar_JCP_1987,
Boehmer_Macromolecules_1990a,
Varoqui_JCP_1991,
Varoqui_JPhysII_1993,
Borisov_JPhysII_1994,
Borukhov_EPL_1995,
Chatellier_JPhysII_1996,
Linse_Macromolecules_1996,
Shubin_Macromolecules_1997,
Borukhov_Macromolecules_1998,
Joanny_EPJB_1999,
Netz_Macromol_1999b,
Dobrynin_PRL_2000,
Dobrynin_Macromolecules_2001,
Dobrynin_JPCB_2003,
Shafir_JCP_2003,
Vries_JCP_2003,
Cheng_JCP_2003,
Lai_condmat_2003,
Lai_condmat_2004}
on the level of mean field theories.
The case of PE adsorption on heterogeneously charged surfaces
was recently theoretically addressed by de Vries et al. \cite{Vries_JCP_2003}.
A remarkable common feature of some of these studies is the charge reversal 
(\textit{overcharging}) of the substrate by the adsorbed PEs 
(see e. g. Refs. \cite{Linse_Macromolecules_1996,
Shubin_Macromolecules_1997,
Joanny_EPJB_1999,Netz_Macromol_1999b,
Grosberg_RevModPhys_2002,
Dobrynin_JPCB_2003}).
The PE adsorption onto \textit{similarly} charged substrates were recently 
investigated by Dobrynin and Rubinstein \cite{Dobrynin_JPCB_2003} and
Cheng and Lai \cite{Lai_condmat_2003,Lai_condmat_2004}.
In the latter situation, the PE adsorption is then driven either by 
\textit{non-electrostatic short range} forces \cite{Dobrynin_JPCB_2003} 
or \textit{attractive image forces} \cite{Lai_condmat_2003,Lai_condmat_2004} 
stemming from a high-dielectric surface.
The problem of \textit{repulsive} image forces stemming from a low-dielectric
surface was studied by Borisov et al. \cite{Borisov_JPhysII_1994} and
Netz and Joanny \cite{Netz_Macromol_1999b} on the level of the
Debye-H\"uckel approximation.

As far as computer simulations are concerned, there exist few Monte Carlo (MC)
studies about PE adsorption on planar charged substrates
\cite{Beltran_Macromolecules_1991,
Kong_JCP_1998,
Yamakov_JPCM_1999,
Ellis_JCP_2000,
Lai_condmat_2003,
Messina_macromol_2004}.
The first MC study on PE adsorption was that of Belt\'an et al. 
\cite{Beltran_Macromolecules_1991} where a lattice-model was employed.
Yamakov et al. \cite{Yamakov_JPCM_1999} performed extensive MC simulations
and found excellent agreement with the scaling predictions of Borisov et al. 
\cite{Borisov_JPhysII_1994}, where different regimes of adsorption are identified. 
Ellis et al.  \cite{Ellis_JCP_2000} 
considered the interesting case of heterogeneously charged surfaces 
(made of positively and negatively charged surface-sites) and demonstrated 
that a PE carrying the same sign of charge as that
of the net charge of the substrate can adsorb. 
Cheng et al. \cite{Lai_condmat_2003} also investigated the effect of image charges
on a high-dielectric constant substrate.
It is to mention that all these MC simulations 
\cite{Beltran_Macromolecules_1991,
Kong_JCP_1998,
Yamakov_JPCM_1999,
Ellis_JCP_2000,
Lai_condmat_2003}
use the Debye-H\"uckel approximation. 
It is only for the case of PE multilayering studied by Messina \cite{Messina_macromol_2004} 
that unscreened long-range electrostatic interactions were properly treated.

In this article, we investigate multi-chain adsorption in the dilute regime but at 
fixed PE volume fraction in a salt-free environment. Counterions from the substrate and the PEs
as well as image forces are explicitly taken into account. 
The influence of chain length (for short chains) and substrate-charge density is
also considered.
Our paper is organized as follows: The model and simulation technique 
are detailed in Sec. \ref{ Sec.simu_method}. Our results are presented in 
Sec. \ref{Sec.Results} and Sec. \ref{Sec.conclu} provides concluding remarks. 

\section{Model and Parameters
 \label{ Sec.simu_method}}

\subsection{Simulation model
\label{ Sec.simu_model}}

The setup of the system under consideration is similar to that recently
investigated with a planar substrate (without image forces) 
\cite{Messina_macromol_2004}. 
Within the framework of the primitive model we consider a PE
solution near a charged hard wall with an implicit solvent 
(water at $z>0$)
of relative dielectric permittivity $\epsilon_{s}\approx 80$.  
The substrate located at $z<0$ is characterized by  
a relative dielectric permittivity $\epsilon_{p}$ which leads
to a dielectric jump $\Delta_{\epsilon}$ 
(when $\epsilon_{s} \neq \epsilon_{p}$) at the interface defined as 
%
\begin{eqnarray}
\label{eq.Delta_eps}
\Delta_{\epsilon}  = 
\frac{\epsilon_{s} - \epsilon_{p}}{\epsilon_{s} + \epsilon_{p}}
\geq 0.
\end{eqnarray}
%

The \textit{negative} bare surface-charge density of the substrate is $-\sigma_0 e$, where $e$ is the
(positive) elementary charge and $\sigma_0>0$ is the number of charges per unit area. 
Electroneutrality is always ensured by the presence of explicit monovalent 
($Z_c=1$) plate's counterions (i.e., monovalent cations) of diameter $a$.
PE chains are made up of $N_m$ \textit{monovalent} positively charged monomers 
($Z_{m}=1$) of diameter $a$.
Their counterions (monovalent anions) are also explicitly taken into account with 
the same parameters up to to the charge sign as the monomers. 
Hence, all microions are monovalent: $Z=Z_c=Z_m=1$ with the same diameter size $a$.

All these particles making up the system are immersed in a rectangular
$L \times L \times  \tau$ box.
Periodic boundary conditions are applied in the $(x,y)$ directions, whereas hard walls
are present at $z=0$ (location of the charged interface) and $z=\tau$ 
(location of an \textit{uncharged} wall).

The total energy of interaction of the system can be written as

\begin{eqnarray}
\label{eq.U_tot}
U_{tot} & = &  
\sum_i  \left[ U_{hs}^{(plate)}(z_i) + U_{coul}^{(plate)}(z_i) \right] + 
\\ \nonumber
&& \sum _{i,i<j} \left[U_{hs}(r_{ij}) + U_{coul}({\bf r}_i, {\bf r}_j) 
                     + U_{FENE}(r_{ij}) + U_{LJ}(r_{ij}) \right],
\end{eqnarray}
where the first (single) sum stems from the interaction between an ion $i$ 
(located at $z=z_i$) and the charged plate, 
and the second (double) sum stems from the pair interaction between ions $i$ and $j$
with $r_{ij}=|{\bf r}_i - {\bf r}_j|$.
All these contributions to $U_{tot}$ in Eq. (\ref{eq.U_tot})
are described in detail below.

Excluded volume interactions are modeled via a hardcore potential
\cite{note_HS} defined as follows
%
\begin{equation}
\label{eq.U_hs}
U_{hs}(r_{ij})=\left\{
\begin{array}{ll}
0,
& \mathrm{for}~r_{ij} \geq a \\
\infty,
& \mathrm{for}~r_{ij} < a 
\end{array}
\right.
\end{equation}
%
for the microion-microion one, and
%
\begin{equation}
\label{eq.U_hs_plate}
U_{hs}^{(plate)}(z_i)=\left\{
\begin{array}{ll}
0,
& \mathrm{for} \quad a/2 \leq z_i \leq  \tau - a/2 \\
\infty,
& \mathrm{otherwise}
\end{array}
\right.
\end{equation}
%
for the plate-microion one.
For clarity, we recall that a microion stands either for a (charged) monomer or a counterion.

The electrostatic energy of interaction between two microions $i$ and $j$ reads
%
\begin{equation}
\label{eq.U_coul} 
\beta U_{coul}({\bf r}_i, {\bf r}_j) =
\pm l_B \left[  
  \frac{1}{r_{ij}} 
+ \frac{\Delta_{\epsilon}}{\sqrt{x_{ij}^2 + y_{ij}^2 + (z_i+z_j)^2}}
\right],
\end{equation}
%
where +(-) applies to microions of the same (opposite) sign, 
$l_{B}=\beta e^{2}/4\pi \epsilon _{0}\epsilon _{r}$ is the Bjerrum
length corresponding to the distance at which two protonic charges
interact with $1/\beta=k_B T$, and $\Delta_{\epsilon}$ is given by Eq.~\eqref{eq.Delta_eps}.  
The first term in Eq.~\eqref{eq.U_coul} corresponds
to the direct Coulomb interaction between real ions,
whereas the second term represents the interaction between the real ion $i$
and the image of ion $j$. By symmetry, the latter also describes the interaction
between the real ion $j$ and the image of ion $i$ yielding
an implicit factor $1/2$.
The electrostatic energy of interaction between an ion $i$ and the
(uniformly) charged plate reads
%
\begin{equation}
\label{eq.U_coul_plate} 
\beta U_{coul}^{(plate)}(z_i) =
l_B \left[ 
  \pm 2 \pi  \sigma_0 z_i
  + \frac{\Delta_{\epsilon}}{4z_i}
\right],
\end{equation}
%
where, for the first term, +(-) applies to positively (negatively) charged ions.
The second term in Eq.\eqref{eq.U_coul_plate} stands for the \textit{self-image} 
interaction, i.e., the interaction between the ion $i$ and its own image.
An appropriate and efficient modified Lekner sum was utilized to compute 
the electrostatic interactions with periodicity in \textit{two} 
directions \cite{Brodka_MolPhys_2002}. 
To link our simulation parameters
to experimental units and room temperature ($T=298$K) we choose
$a =4.25$ \AA\ leading to the Bjerrum length of water
$l_{B}=1.68a =7.14$ \AA. In order to investigate  the effect of
image forces we take a value of $\epsilon_p=2$ for the
dielectric constant of the charged substrate 
(which is a typical value for silica or mica substrates \cite{Malinsky_JPCB_2001}) 
and $\epsilon_s=80$ for that of the aqueous solvent yielding
$\Delta_{\epsilon}=\frac{80 - 2}{80+2} \approx 0.951$.
The case of identical dielectric constants $\epsilon_p=\epsilon_s$
($\Delta_{\epsilon}=0$) corresponds to the situation where there are no
image charges.

The  PE chain connectivity is modeled by employing a standard
finite extendible nonlinear elastic (FENE) potential for good solvent,
which reads
%
\begin{equation}
\label{eq.U_fene}
U_{FENE}(r)=
\left\{ \begin{array}{ll}
\displaystyle -\frac{1}{2}\kappa R^{2}_{0}\ln \left[ 1-\frac{r^{2}}{R_{0}^{2}}\right] ,
& \textrm{for} \quad r < R_0 \\ \\
\displaystyle \infty ,
& \textrm{for} \quad r \geq R_0 \\
\end{array}
\right.
\end{equation}
%
with $\kappa = 27k_{B}T/ a^2$ and $R_{0}=1.5 a$.
The excluded volume interaction between chain monomers is taken into
account via a shifted and truncated Lennard-Jones (LJ) potential given
by
%
\begin{equation}
\label{eq.LJ}
U_{LJ}(r)=
\left\{ \begin{array}{ll}
\displaystyle
4\epsilon \left[ \left(\frac{a}{r}\right)^{12}
-\left( \frac{a}{r}\right) ^{6}\right] +\epsilon,
& \textrm{for} \quad r \leq 2^{1/6} a \\ \\
0,
& \textrm{for} \quad  r > 2^{1/6} a
\end{array}
\right.
\end{equation}
%
where $\epsilon=k_BT$.
These parameter values lead to an equilibrium bond length $ l=0.98a$.

%
\begin{table}[b]
\caption{
List of key parameters with some fixed values.
}
\label{tab.simu-param}
\begin{ruledtabular}
\begin{tabular}{lc}
 Parameters&
\\
\hline
 $T=298K$&
 room temperature\\
 $\sigma_0 L^2$&
 charge number of the substrate\\
 $\Delta_{\epsilon} = 0 ~ {\rm or} ~ 0.951$ &
 dielectric discontinuity\\
 $Z=1$&
 microion valence\\
 $a =4.25$ \AA\ &
 microion diameter\\
 $l_{B}=1.68a =7.14$ \AA\ &
 Bjerrum length\\
 $L=25 a $&
 $(x,y)$-box length\\
 $\tau=75 a $&
 $z$-box length\\
 $N_{PE}$&
 number of PEs\\
 $N_m$&
 number of monomers per chain
\end{tabular}
\end{ruledtabular}
\end{table}
%

All the simulation parameters are gathered in Table \ref{tab.simu-param}.
The set of simulated systems can be found in Table \ref{tab.simu-runs}.
The equilibrium properties of our model system were obtained by using standard canonical MC 
simulations following the Metropolis scheme \cite{Metropolis_JCP_1953,Allen_book_1987}.
Single-particle moves were considered with an acceptance ratio of
$30\%$ for the monomers and $50\%$ for the counterions.
Depending on the parameters, the length of a simulation run ranges
from $2\times 10^6$ up to $7\times 10^6$  MC steps per
particle.
Typically, about $3 \times 10^5$ to $2.5 \times 10^6$ MC steps 
were required for equilibration, and 
$1-4 \times 10^6$ subsequent MC steps were used to perform measurements. 

\begin{table}
\caption{
Simulated systems' parameters. The number of counterions (cations and anions) ensuring
the overall electroneutrality of the system is not indicated.
}
\label{tab.simu-runs}
\begin{ruledtabular}
\begin{tabular}{lccc}
 System&
 $N_{PE}$&
 $N_m$&
 $\sigma_0L^2$
\\
\hline
 $A$&
 $96$&
 $2$&
 $64$\\
 $B$&
 $48$&
 $4$&
 $64$\\
 $C$&
 $24$&
 $8$&
 $64$\\
 $D$&
 $12$&
 $16$&
 $64$\\
 $E$&
 $6$&
 $32$&
 $64$\\
 $F$&
 $12$&
 $16$&
 $32$\\
$G$&
 $12$&
 $16$&
 $128$\\
$H$&
 $12$&
 $16$&
 $192$\\
\end{tabular}
\end{ruledtabular}
\end{table}

\subsection{Measured quantities
 \label{Sec.Target}}

We briefly describe the different observables that are going to be measured.  
In order to study the PE adsorption, we compute the monomer density
$n(z)$ that is normalized as follows

\begin{equation}
\label{eq.n_z}
\int ^{\tau-a/2}_{a/2} n(z) L^2 dz = N_{PE} N_m.
\end{equation}
%
%
To further characterize the PE adsorption, we also compute the total number of  
accumulated monomers $\bar{N}(z)$ within a distance $z$
from the planar charged plate that is given by
%
\begin{equation}
\label{eq.N_z}
\bar{N}(z) = \int ^{z}_{a/2} n(z') L^2 dz'.
\end{equation}
%
It is useful to introduce the fraction of adsorbed monomers,
$N^*(z)$, which is defined as follows
%
\begin{equation}
\label{eq.N_z_star}
N^*(z) = \frac{\bar{N}(z)}{N_{PE} N_m}.
\end{equation}
%

Another relevant quantity is the global \textit{net fluid
charge} $\sigma(z)$ which is defined as follows
\begin{equation}
\label{Eq.Qz}
\sigma(z)=\int ^{z}_{a/2} \left[
n_{+}(z') - n_{-}(z')\right] dz',
\end{equation}
%
where $n_+$ and $n_-$ stand for the density of all the
positive microions (i.e., monomers and plate's counterions) and
negative microions (i.e., PEs' counterions), respectively.
It is useful to introduce the reduced surface charge density
$\sigma^*(z)$ defined as follows:
\begin{equation}
\label{Eq.Qz_star}
\sigma^*(z) = \frac{\sigma(z)}{\sigma_0}.
\end{equation}
%
Thereby, $\sigma^*(z)$ corresponds, up to a prefactor $\sigma_0 e$, 
to the net fluid charge per unit area 
(omitting the surface charge density $-\sigma_0e$ of the substrate)
within a distance $z$ from the charged wall. 
At the uncharged wall, electroneutrality imposes $\sigma^*(z=\tau-a/2)=1$.  
By simple application of the Gauss' law,
$\left[ \sigma^*(z)-1\right]$ is directly proportional
to the mean electric field at $z$.  Therefore $\sigma^*(z)$ can
measure the \textit{screening} strength of the substrate
by the neighboring solute charged species.

\section{Results and discussion
 \label{Sec.Results}}

From previous studies 
\cite{Torrie_JCP_1982,Messina_image_2002,Borisov_JPhysII_1994,Netz_Macromol_1999b} 
it is well understood that effects of image charges become especially
relevant at sufficiently low surface charge density of the interface. 
It is also clear that the self-image interaction 
(\textit{repulsive} for $\Delta_{\epsilon}>0$, as is presently the case)
is higher the higher the charge of the ions (polyions) since it 
scales like $Z^2$. In the present situation where we have to deal with 
PEs, the length of the chain ($N_m$) is a key parameter that can be seen as
the valence of a polyion.
Hence, we are going to study (i) the influence of chain length 
(Sec. \ref{Sec.chain_length}) and (ii) that of surface charge density
(Sec. \ref{Sec.charge_density}).
For the sake of consistency, we fixed the total number of monomers
to $N_{PE}N_m=192$ meaning that the monomer concentration is 
\textit{fixed} (see also Table \ref{tab.simu-runs}): The PE volume fraction 
$\phi = \frac{4\pi}{3}\frac{N_{PE}N_m(a/2)^3}{L^2\tau} \approx 2.14 \times 10^{-3}$ 
is fixed.

\subsection{Influence of chain length
 \label{Sec.chain_length}}

In this part, we consider the influence of chain length $N_m$
at fixed surface charge density parameter $\sigma_0L^2=64$.
The latter corresponds experimentally to a moderate \cite{Tulpar_JPCB_2004}
surface charge density with $-\sigma_0e \approx -0.091 ~ {\rm C/m^2}$.
The chain length is varied from $N_m=2$ up to $N_m=32$ 
(systems $A-E$, see Table \ref{tab.simu-runs}).
We have ensured that, for the longest chain with $N_m=32$,
size effects are still negligible since the end-to-end distance
is about $8.8a$ which is significantly smaller than $L=25a$ or $\tau=75a$.

\begin{figure}
\includegraphics[width = 8.0 cm]{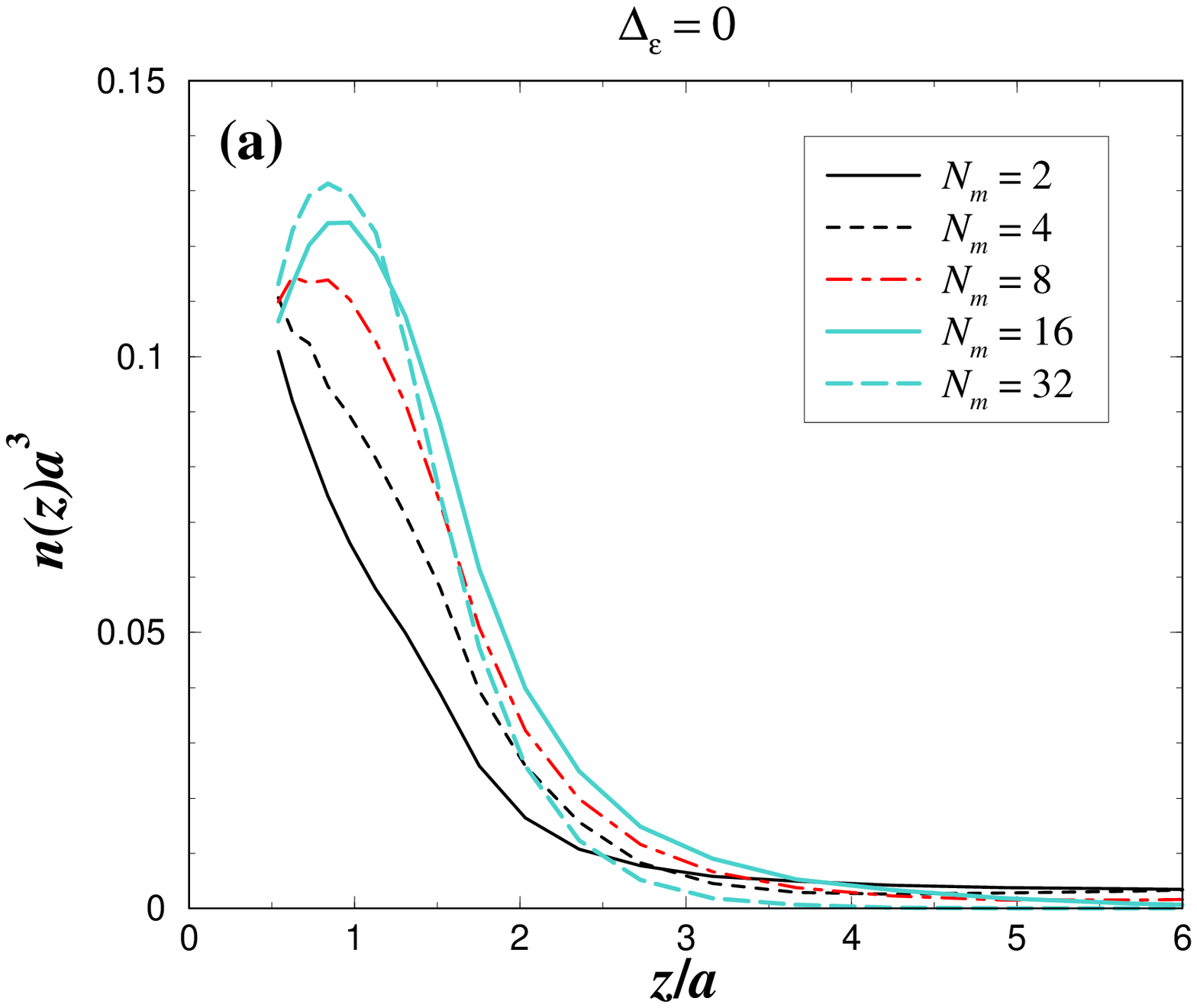}
\includegraphics[width = 8.0 cm]{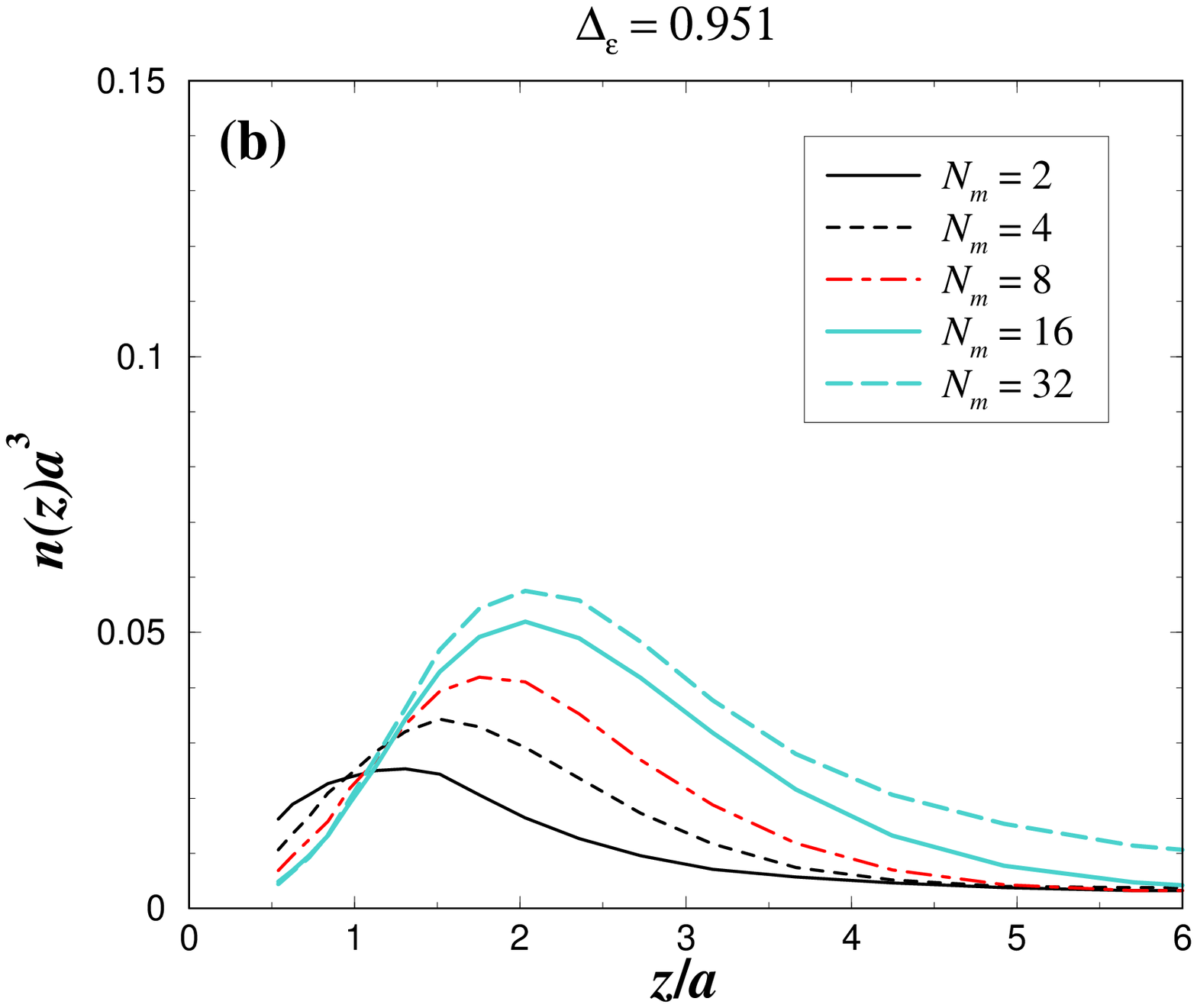}
\caption{
Profiles of the monomer density $n(z)$ for different chain length $N_m$ 
with $\sigma_0L^2=64$ (systems $A-E$).
(a) $\Delta_{\epsilon}=0$.
(b) $\Delta_{\epsilon}=0.951$.
}
\label{fig.nz_Qp-64}
\end{figure}

\begin{figure}
\includegraphics[width = 8.0 cm]{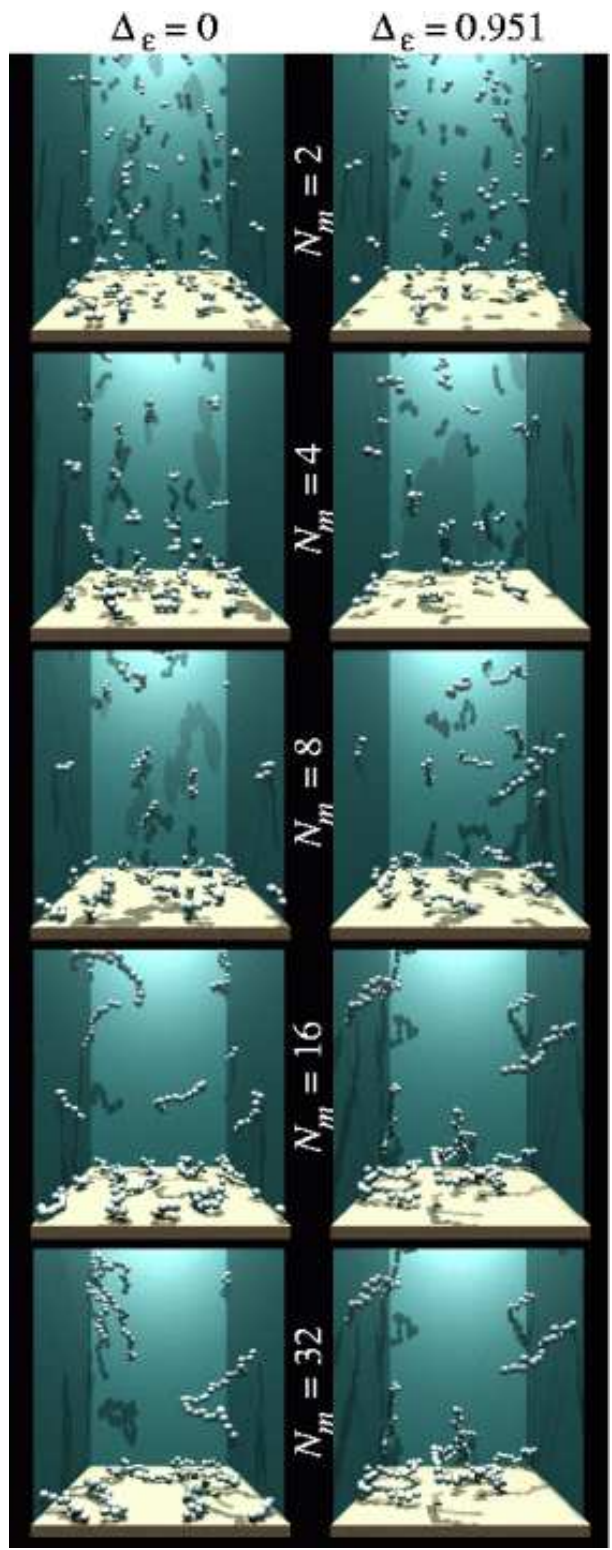}
\caption{
Typical equilibrium microstructures of systems $A-E$.
The little counterions are omitted for clarity.
}
\label{fig.snap_Qp-64}
\end{figure}

The profiles of the monomer distribution $n(z)$ can be found in
Fig. \ref{fig.nz_Qp-64} and the corresponding microstructures are sketched in 
Fig. \ref{fig.snap_Qp-64}. 
Let us first comment the more simple case where no image charges 
are present [$\Delta_{\epsilon}=0$ - Fig. \ref{fig.nz_Qp-64}(a)].
For (very) short chains (here $N_m \leq 4$), Fig. \ref{fig.nz_Qp-64}(a) shows that
the density profiles exhibit a monotonic behavior \textit{even near contact}.
Within this regime of chain length, the monomer density near the 
charged wall increases with increasing $N_m$. This feature is 
fully consistent with the idea that stronger $lateral$ correlations,
the latter scaling like $Z^{3/2}$ for spherical counterions at 
fixed $\sigma_0$ \cite{Shklowskii_PRE_1999b,Messina_PRE_2001},
induce a higher polyion adsorption. In other words, at (very) low $N_m$ 
\textit{conformational entropic} effects are not dominant and the short-chains 
systems can be qualitatively understood with the picture provided
by spherical (or point-like) ions.
The scenario becomes qualitatively different at higher chain length
[here $N_m \geq 8$ - see Fig. \ref{fig.nz_Qp-64}(a)], where
$n(z)$ presents a maximum near contact which is the signature of a 
\textit{short range repulsion} that was also theoretically predicted 
by Borukhov et al. \cite{Borukhov_EPL_1995}.
This non-trivial feature can be explained in terms of entropy:
Near the surface of the substrate the number of available PE conformations 
is considerably reduced yielding to an entropic repulsion that can be detected 
if the driving force of PE adsorption (crucially controlled by $\sigma_0$) 
is not strong enough.
This latter statement will be properly examined and confirmed in 
Sec. \ref{Sec.charge_density}
where the influence of $\sigma_0$ is addressed.
Nonetheless, the highest value of $n(z;N_m)$ increases with $N_m$ as it should be.
All these mentioned features can be visualized on the microstructures 
depicted in Fig. \ref{fig.snap_Qp-64}.
%
One can summarize those relevant findings, valid for small enough  
$\sigma_0$ and $\Delta_{\epsilon}=0$, as follows:
\begin{itemize}
\item For very short chains the PE adsorption is similar to that occurring
      with spherical electrolytes.
\item PE chains experience a short range repulsion near the substrate due to conformational 
      entropic effects.
\end{itemize}

%

We now turn to the more complicated situation where image forces are present 
[$\Delta_{\epsilon}=0.951$ - Fig. \ref{fig.nz_Qp-64}(b)].
An immediate remark that can be drawn from a comparison with the
$\Delta_{\epsilon}=0$ case is that the PE adsorption is much weaker
due to the repulsive image-polyion interactions. 
At all $N_m$, $n(z)$ presents a maximum at $z=z^*$ that is
gradually shifted to larger $z$ with increasing $N_m$. 
In other words, the \textit{thickness} of the adsorbed PE layer as determined by $z^*$ 
increases with $N_m$.
This phenomenon is of course due to the fact that the
image-polyion repulsion increases with $N_m$, similarly
to what happens with multivalent (point-like or spherical) counterions 
\cite{Torrie_JCP_1982,Messina_image_2002}.
On the other hand, interestingly, the monomer density at contact 
\textit{decreases} with increasing $N_m$. 
This is the result of a \textit{combined} effect of (i) conformational entropy as 
explained above and (ii) the $N_m$-induced image-polyion repulsion. 
All those features are well illustrated on the microstructures of 
Fig. \ref{fig.snap_Qp-64}.

%

\begin{figure}
\includegraphics[width = 8.0 cm]{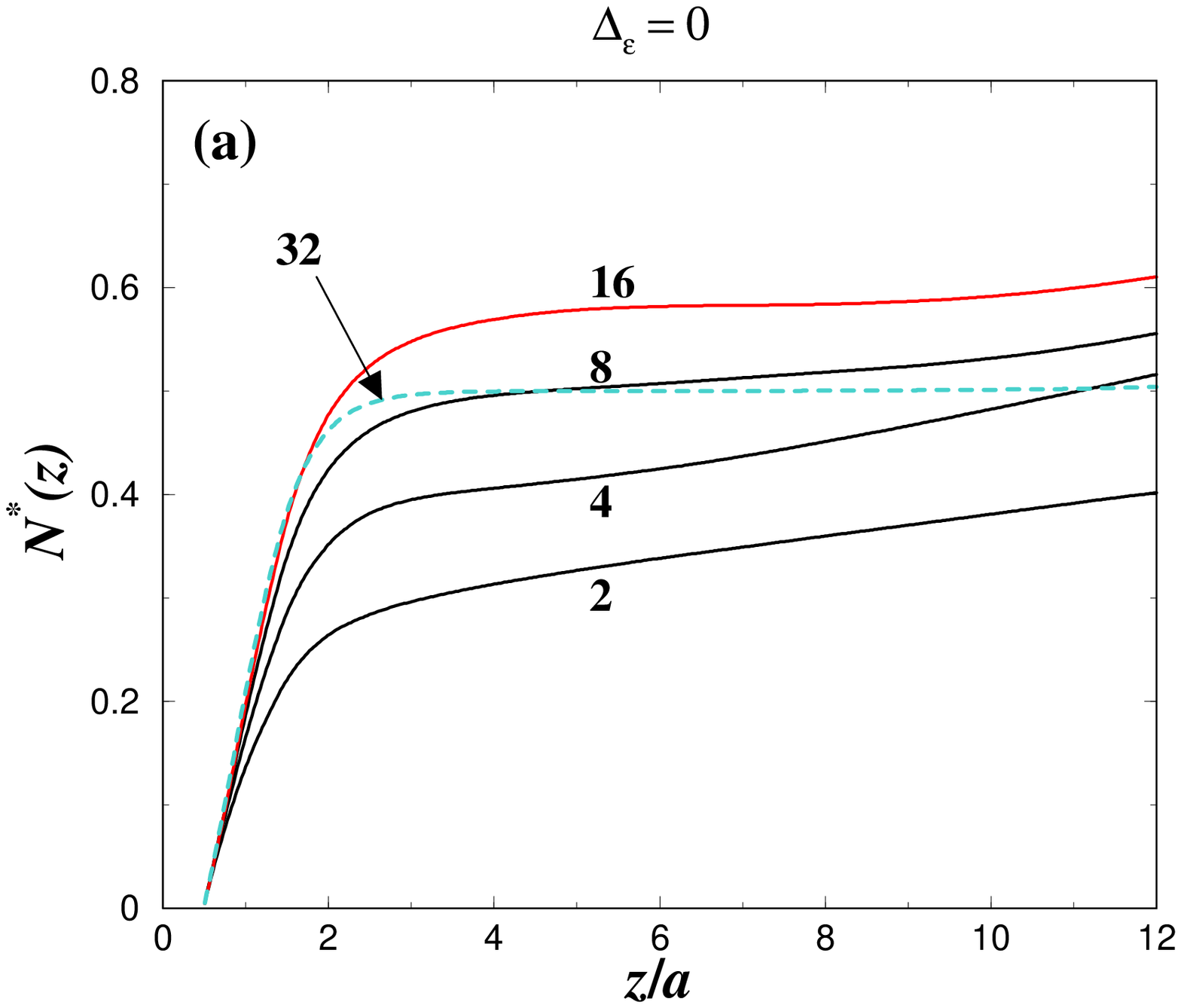}
\includegraphics[width = 8.0 cm]{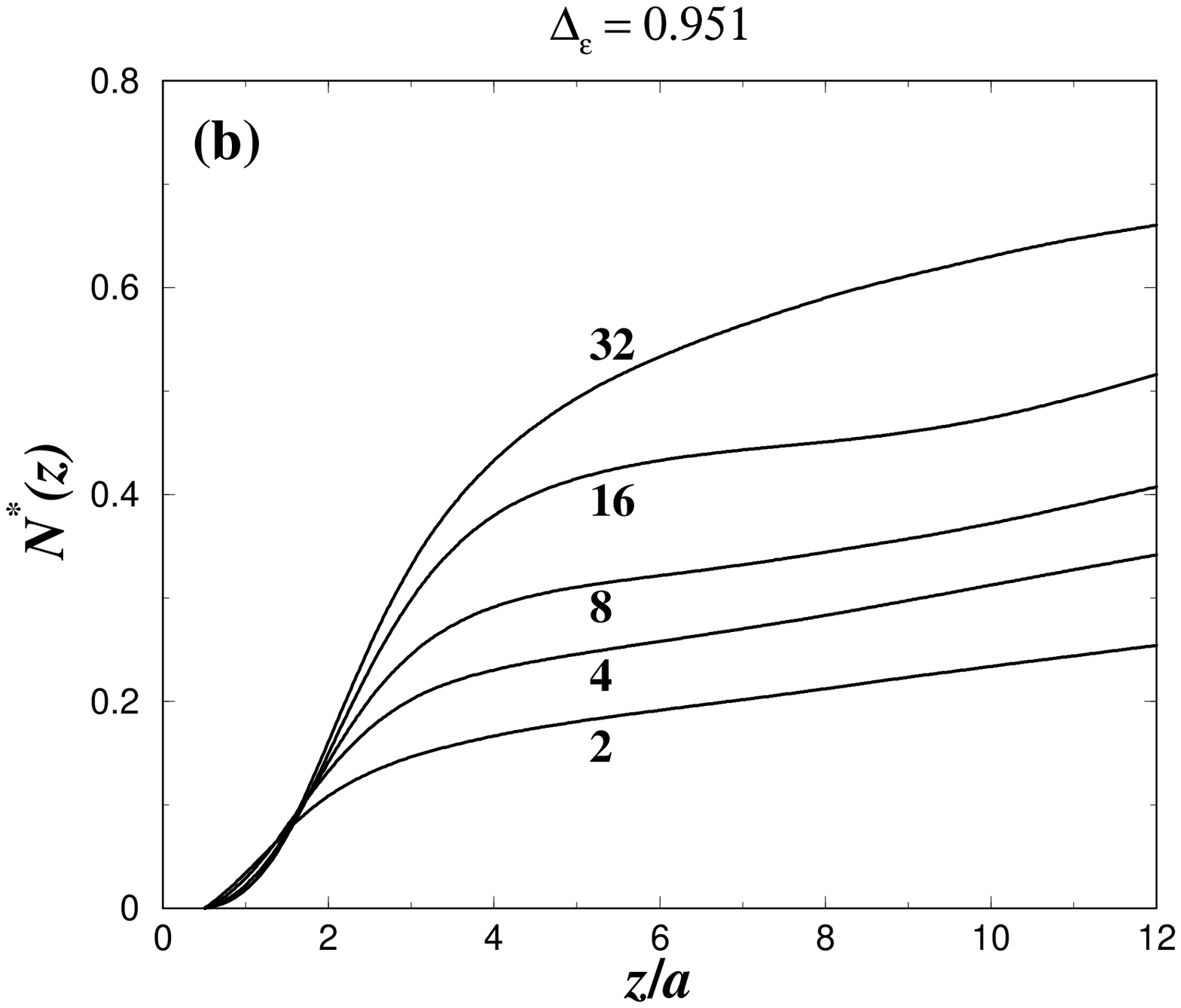}
\caption{
Profiles of the fraction of adsorbed monomers $N^*(z)$ 
for different chain length $N_m$ (as indicated by its numerical value) 
with $\sigma_0L^2=64$ (systems $A-E$).
(a) $\Delta_{\epsilon}=0$.
(b) $\Delta_{\epsilon}=0.951$.
}
\label{fig.Nz_star_Qp-64}
\end{figure}

To gain further insight into the properties of PE adsorption, we
have plotted the fraction of adsorbed monomers $N^*(z)$ 
[Eq.~\eqref{eq.N_z_star}] in Fig. \ref{fig.Nz_star_Qp-64}.
At $\Delta_{\epsilon}=0$ [Fig. \ref{fig.Nz_star_Qp-64}(a)], 
it is observed in the immediate vicinity of the wall 
(roughly for $z \lesssim 1.5a$) that $N^*(z;N_m)$ increases monotonically 
with $N_m$ as expected.
However, further away from the wall, a non-trivial effect is found where $N^*(z;N_m)$ 
surprisingly exhibits a non-monotonic behavior with respect to $N_m$. 
More explicitly, in the regime of large $N_m$ we have
$N^*(z;N_m=32)$ that is clearly smaller than $N^*(z;N_m=16)$
and even smaller than $N^*(z;N_m=8)$ when one is sufficiently far from the wall.
This remarkable phenomenon is going to be explained later by advocating
the role of overcharging. 
Upon switching the image forces on
[$\Delta_{\epsilon}=0.951$ - Fig. \ref{fig.Nz_star_Qp-64}(b)], 
$N^*(z;N_m)$ shows a qualitatively different behavior than that
found at $\Delta_{\epsilon}=0$, in accordance with our study
concerning $n(z)$. More precisely:
(i) very close to the wall, $N^*(z;N_m)$ \textit{decreases} with $N_m$
whilst (ii) sufficiently far away from the wall $N^*(z;N_m)$ \textit{increases} with $N_m$. 
This behavior is fully consistent with our mechanisms previously discussed for $n(z)$. 
Below, we are going to show that the reduced net fluid charge $\sigma^*(z)$ 
is a key observable to account for those reported properties of $N^*(z;N_m)$.

%

\begin{figure}
\includegraphics[width = 8.0 cm]{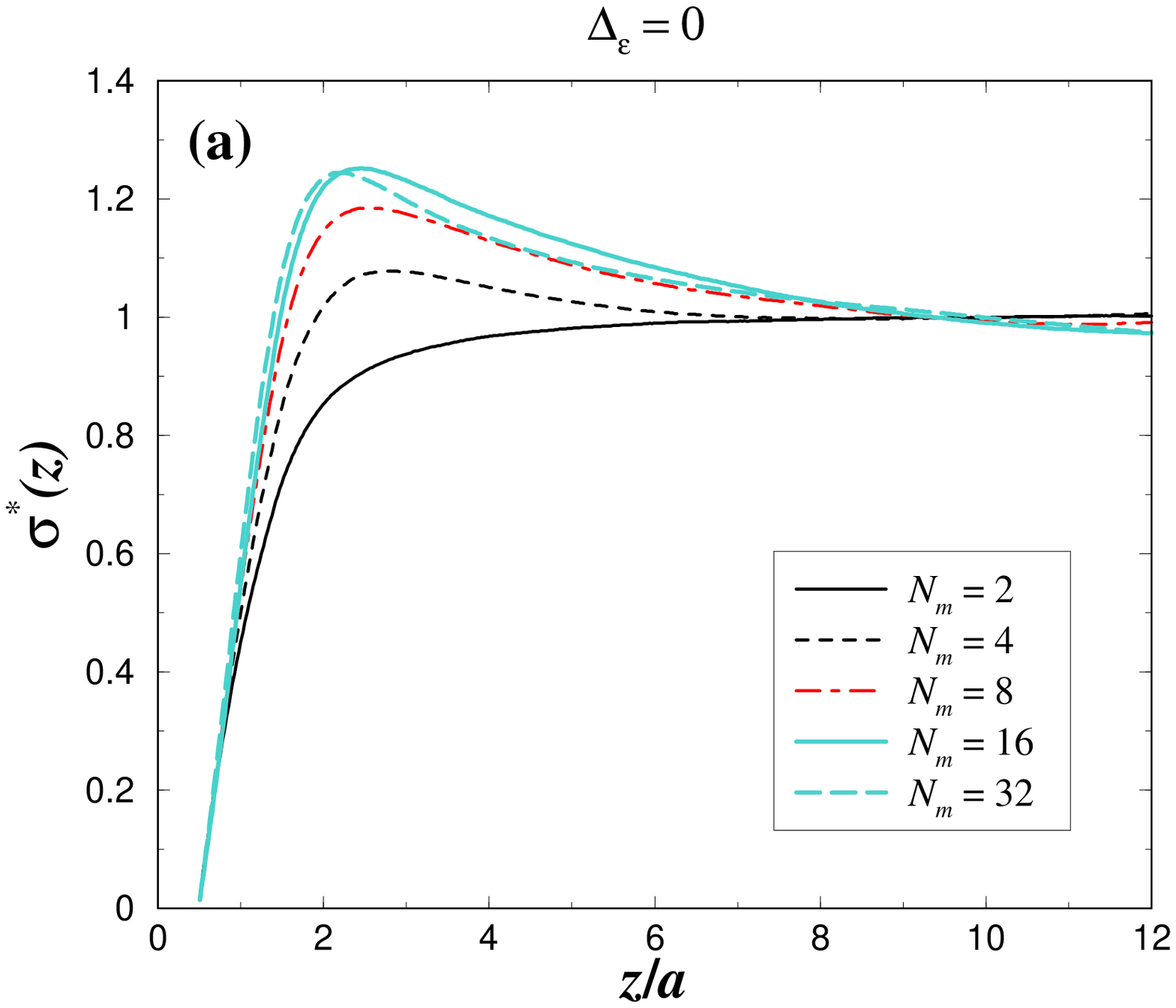}
\includegraphics[width = 8.0 cm]{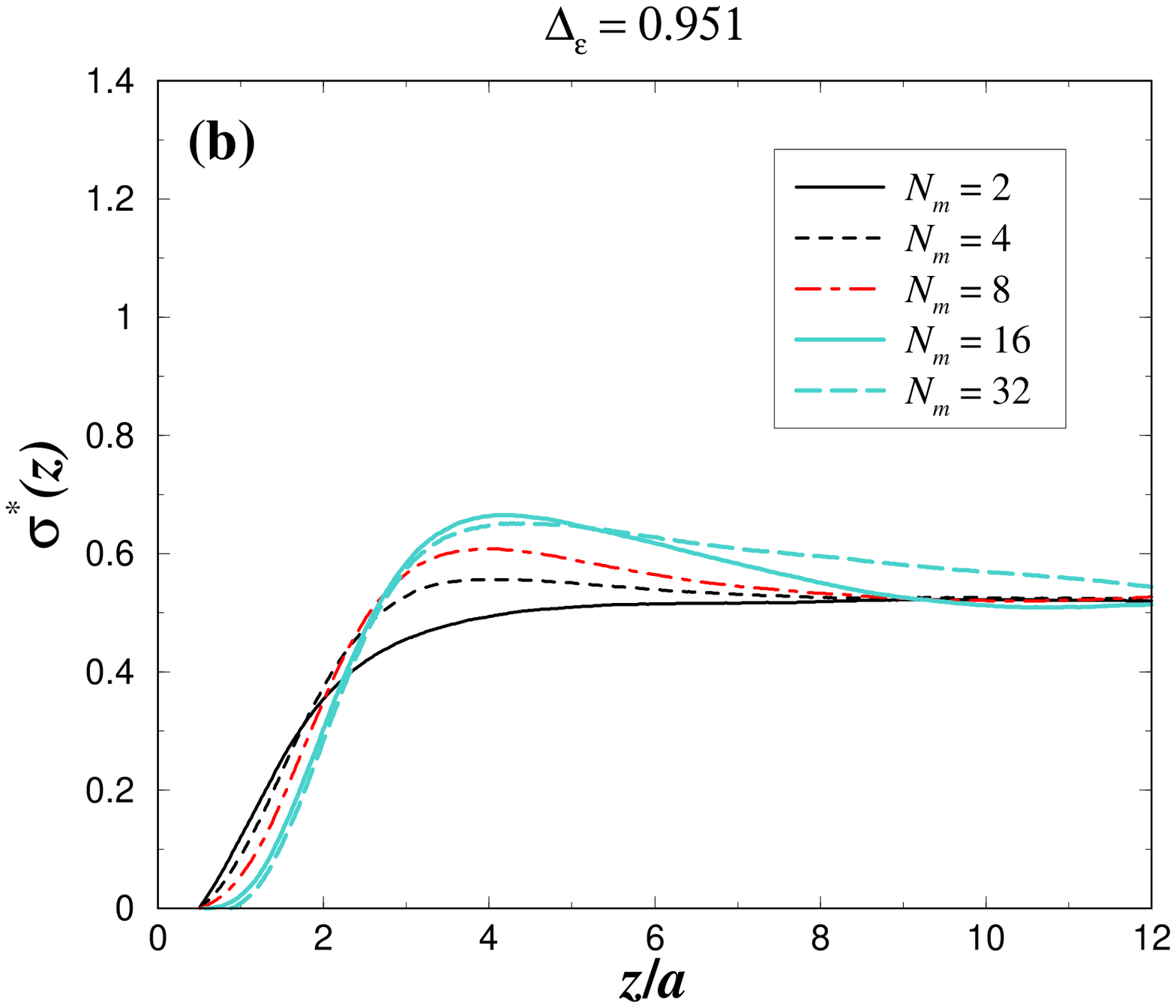}
\caption{
Profiles of the reduced net fluid charge $\sigma^*(z)$ 
for different chain length $N_m$ with $\sigma_0L^2=64$ (systems $A-E$).
(a) $\Delta_{\epsilon}=0$.
(b) $\Delta_{\epsilon}=0.951$.
}
\label{fig.Qz_star_Qp-64}
\end{figure}

A deeper understanding of the physical mechanisms involved in PE adsorption can 
be gained by considering the net fluid charge parameter $\sigma^*(z)$ 
[Eq.~\eqref{Eq.Qz_star}] that describes the screening of the charged interface.
The profiles of $\sigma^*(z)$ for different $N_m$ can be found in 
Fig. \ref{fig.Qz_star_Qp-64}.
At $\Delta_{\epsilon}=0$ [see Fig. \ref{fig.Qz_star_Qp-64}(a)], it is shown that
for long enough chains (here $N_m \geq 4$) the substrate gets 
locally \textit{overcharged} as signaled by $\sigma^*(z)>1$. 
Physically, this means that the global local charge of the adsorbed monomers 
\cite{note_overscreening} is larger in absolute value than that of the plate's charge. 
In other words, the plate is \textit{overscreened} by the adsorbed PE chains. 
Fig. \ref{fig.Qz_star_Qp-64}(a) indicates that the degree of overcharging increases 
with $N_m$ as expected from the behavior of multivalent counterions, and seems
to saturate at high $N_m$.
This enhanced $N_m$-overcharging leads to a sufficiently strong effective 
repulsion between the substrate and the PEs in the solution, which in turn
prevents from further adsorption. It is precisely this mechanism that
explains the apparent anomaly found in Fig. \ref{fig.Nz_star_Qp-64}(a)
where, sufficiently away from the surface, it was reported a significantly 
lower monomer fraction $N^*(z;N_m)$ at $N_m=32$ than at $N_m=16$ or $N_m=8$.   
This spectacular effect is well illustrated in Fig. \ref{fig.snap_Qp-64}
(with $N_m=32$) where, above the (strongly bound) adsorbed PEs, there is a 
depletion region leading to a plateau in $N^*(z;N_m=32)$.

%

Upon inducing polarization charges 
[$\Delta_{\epsilon}=0.951$ - Fig. \ref{fig.Qz_star_Qp-64}(b)]
overscreening is canceled.
This, in turn, accounts for the absence of plateau
in $N^*(z;N_m)$ at $\Delta_{\epsilon}=0.951$. 
That striking disappearance of overcharging can be rationalized by establishing
again an analogy with multivalent spherical ions:
\begin{itemize}
\item For the sake of simplicity, let us assume that the PE can be 
      electrostatically envisioned as a spherical polyion of valence $N_m$
      with a radius corresponding roughly to the radius of gyration of the chain.    
      Thereby, the image-polyion \textit{repulsive} interactions 
      [including the self-image repulsion as well as the lateral image-ion correlations 
      as given by the second term of Eq. \eqref{eq.U_coul}] 
      scale like $N_m^2$ whereas the \textit{attractive} driving force 
      of polyion adsorption due to Wigner crystal ordering scales like $N_m^{3/2}$
      \cite{Messina_image_2002}.
      The latter driving force corresponds to the highest possible attraction between
      the substrate and the polyion and is therefore a good candidate for the 
      present discussion. Consequently, at large enough $N_m$ image forces will be
      dominant and inhibit overcharging. 
\end{itemize}
%
 
\subsection{Influence of substrate surface-charge density
 \label{Sec.charge_density}}

To complete our investigation, we would like to address the
influence of the substrate charge density on the PE
adsorption in presence of image forces.
In this respect, we consider (at fixed $N_m=16$) three additional 
values of the charge density:
$\sigma_0L^2=32,128,192$ corresponding to the systems
$F,G,H$, respectively (see Table \ref{tab.simu-runs}).

\begin{figure}[b]
\includegraphics[width = 8.0 cm]{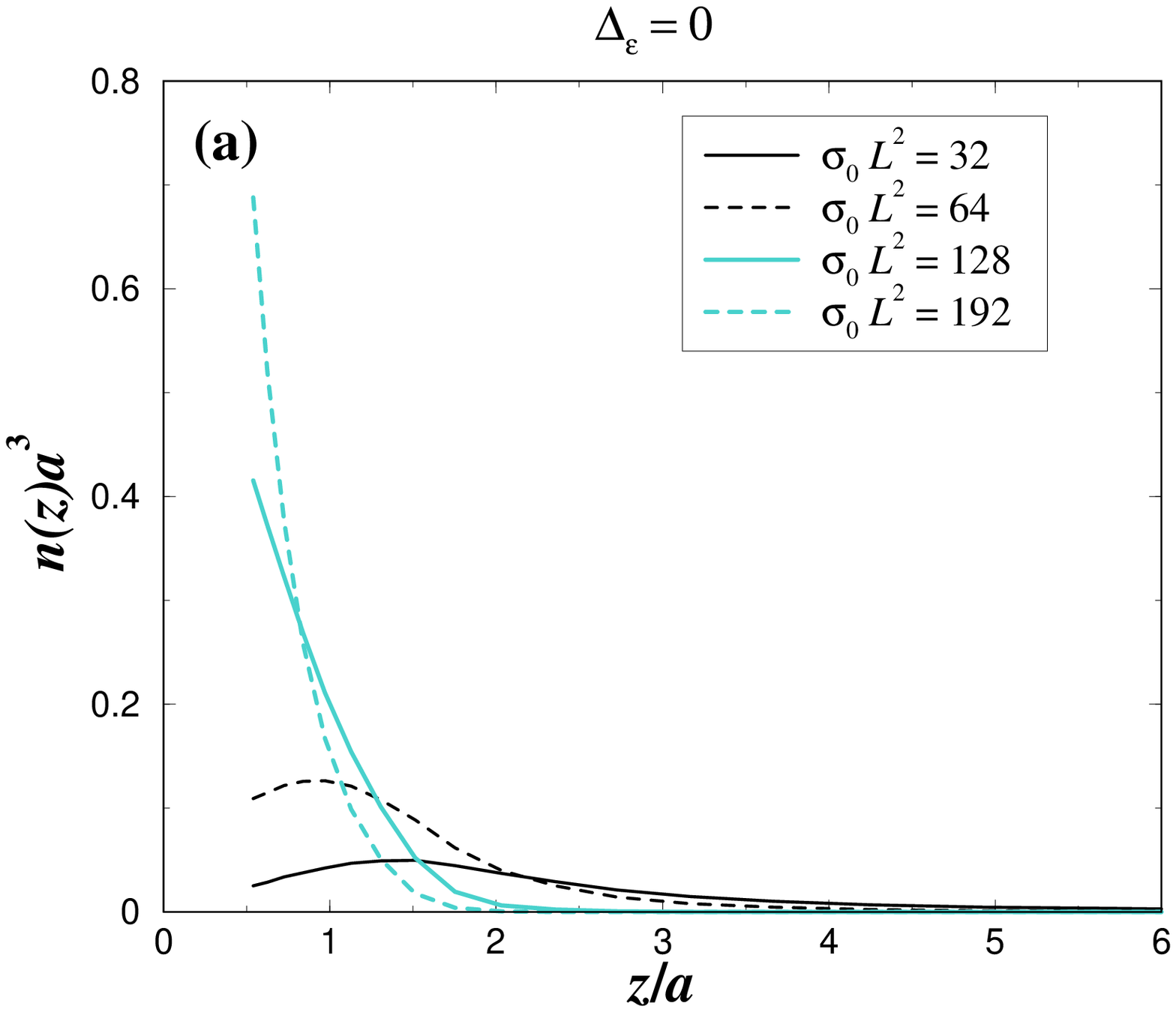}
\includegraphics[width = 8.0 cm]{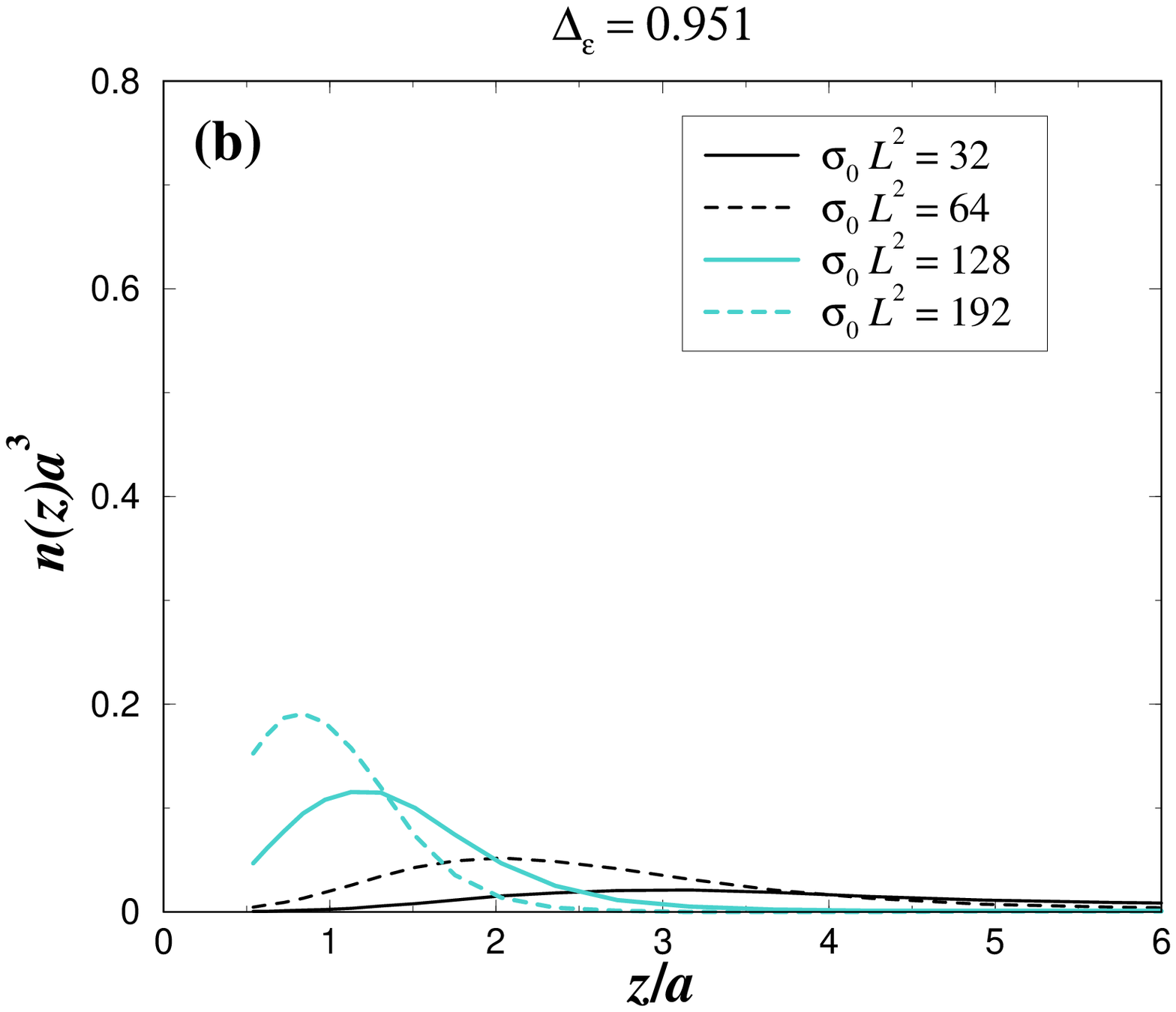}
\caption{
Profiles of the monomer density $n(z)$ for different parameters of surface charge 
density $\sigma_0L^2$ with $N_m=16$ (systems $D,F-H$). 
The case $\sigma_0L^2=64$ (system $D$) from Fig. \ref{fig.nz_Qp-64} 
is reported here again for easier comparison.
(a) $\Delta_{\epsilon}=0$.
(b) $\Delta_{\epsilon}=0.951$.
}
\label{fig.nz_Nm-16}
\end{figure}
%

\begin{figure}
\includegraphics[width = 8.0 cm]{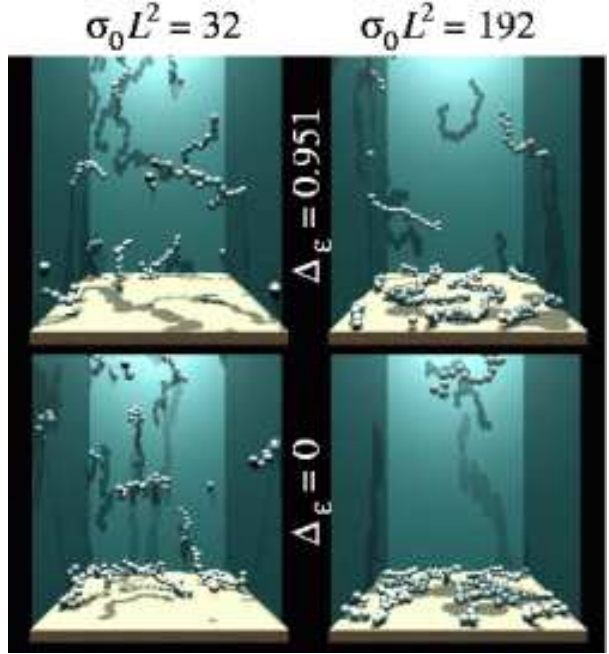}
\caption{
Typical equilibrium microstructures of systems $F$ and $H$.
The little counterions are omitted for clarity.
}
\label{fig.snap_Nm-16}
\end{figure}
%

The plots of the monomer density $n(z)$ at various values 
of $\sigma_0L^2$ can be found in Fig. \ref{fig.nz_Nm-16}.
Microstructures of systems $F$ and $H$ are presented in
Fig. \ref{fig.snap_Nm-16}.
At $\Delta_{\epsilon}=0$ [see Fig. \ref{fig.nz_Nm-16}(a)],
the monomer density at contact increases with $\sigma_0$ as it should be.
Interestingly, the local maximum in $n(z)$ [present at small 
$\sigma_0$ (here $\sigma_0 L^2 \leq 64$)] \textit{vanishes at large} $\sigma_0$
[see Fig. \ref{fig.nz_Nm-16}(a)].
This feature is the result of a $\sigma_0$-enhanced driving force of 
adsorption that overcomes entropic effects at large enough $\sigma_0$.
The strong adsorption at $\sigma_0 L^2 = 192$ leads to a \textit{flat}
PE layer as well illustrated in Fig. \ref{fig.snap_Nm-16}.

By polarizing the substrate surface ($\Delta_{\epsilon}=0.951$), it can be
seen from Fig. \ref{fig.nz_Nm-16}(b) and the snapshot of Fig. \ref{fig.snap_Nm-16}
that there is a strong monomer depletion near contact at $\sigma_0 L^2 = 32$. 
This feature is due to the combined effects of (i) conformational entropy, 
(ii) image-monomer repulsion and (iii) a lower electrostatic wall-monomer attraction.
Upon increasing $\sigma_0$ the monomer density near contact becomes larger, 
and concomitantly, the maximum in $n(z)$ is systematically shifted to smaller $z$.
It is to say that the thickness of the adsorbed PE layer decreases with $\sigma_0$.

\begin{figure}
\includegraphics[width = 8.0 cm]{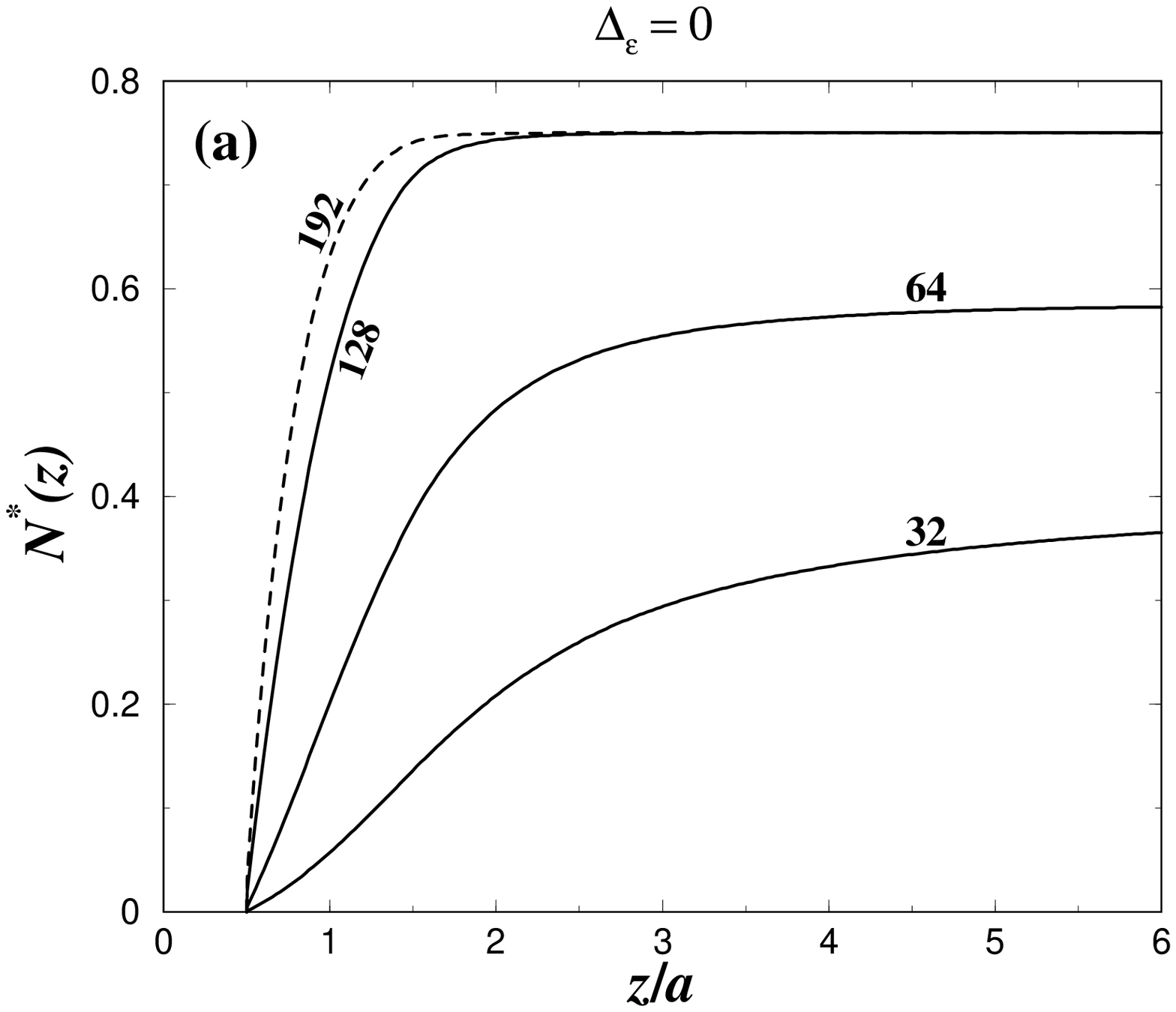}
\includegraphics[width = 8.0 cm]{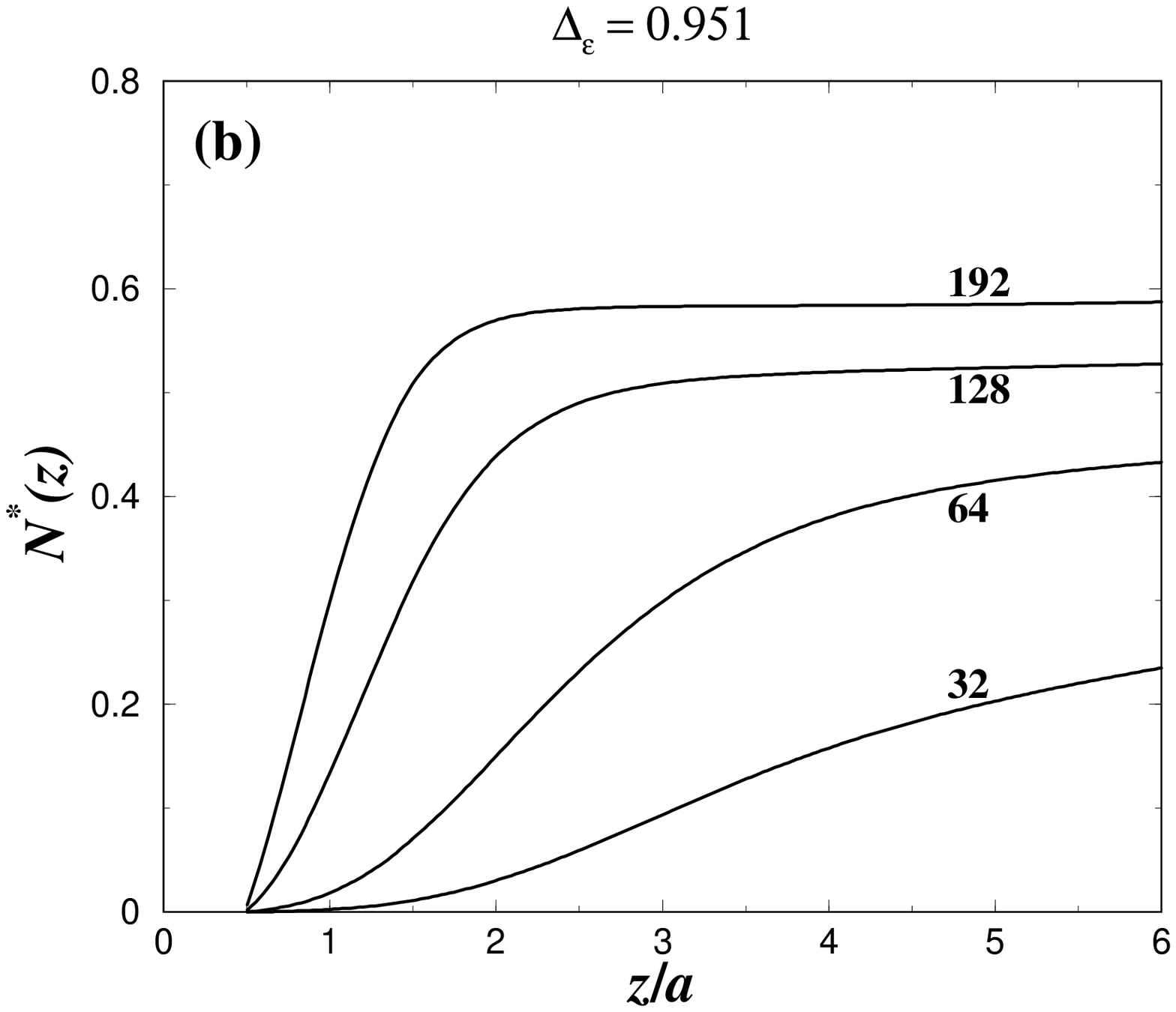}
\caption{
Profiles of the fraction of adsorbed monomers $N^*(z)$ 
for different  parameters of surface charge density $\sigma_0 L^2$ 
(as indicated by its numerical value) with $N_m=16$ (systems $D,F-H$).
The case $\sigma_0L^2=64$ (system $D$) from Fig. \ref{fig.Nz_star_Qp-64}
is reported here again for easier comparison.
(a) $\Delta_{\epsilon}=0$.
(b) $\Delta_{\epsilon}=0.951$.
}
\label{fig.Nz_star_Nm-16}
\end{figure}

The profiles of $N^*(z)$ are provided in Fig. \ref{fig.Nz_star_Nm-16} from which further 
characterization of PE adsorption can be obtained.
At $\Delta_{\epsilon}=0$, Fig. \ref{fig.Nz_star_Nm-16}(a) indicates that 
$N^*(z;\sigma_0)$ increases with $\sigma_0$ but saturates at high $\sigma_0$. 
This latter saturation effect should only be relevant for a regime of charge
where $\eta \equiv \frac{N_{PE}N_m}{\sigma_0L^2}$ is about unity. Indeed, in a
typical experimental situation at finite monomer concentration 
(even in the dilute regime) we have $\eta \ggg 1$ so that overcharging is always possible
at large $\sigma_0$ and thereby $N^*(z;\sigma_0)$ should always significantly increase
with  $\sigma_0$ as long as packing effects 
(as generated by the excluded volume of the monomers) are not vivid.
In parallel, the plateau reported at $\sigma_0L^2=128$ and $\sigma_0L^2=192$
in Fig. \ref{fig.Nz_star_Nm-16}(a), is the signature of a monomer depletion  
above the adsorbed PE layer (see also Fig. \ref{fig.snap_Nm-16}) due to a strong 
screening of the surface charge by the latter. 
At $\Delta_{\epsilon}=0.951$, Fig. \ref{fig.Nz_star_Nm-16}(b) shows that $N^*(z)$ is 
considerably smaller than at $\Delta_{\epsilon}=0$ even for high $\sigma_0$,
in accordance with the behavior of $n(z)$ from Fig. \ref{fig.nz_Nm-16}.
The $\Delta_{\epsilon}$-induced desorption is especially strong at 
$\sigma_0L^2=32$ where the image-monomer repulsion clearly counterbalances 
the electrostatic wall-monomer  attraction. More quantitatively,
at $z=3a$ (a $z$-distance corresponding roughly to the radius of gyration of the chain
with $N_m=16$) about $30\%$ [i.e., $N^*(z)=0.3$] of the monomers are adsorbed  
with $\Delta_{\epsilon}=0$ against only $10\%$ with $\Delta_{\epsilon}=0.951$
[see Fig. \ref{fig.Nz_star_Nm-16}(b)].

%

\section{Concluding remarks
 \label{Sec.conclu}}

We first would like to make some final remarks about the presented results.
As far as the charge surface distribution on the substrate's surface is concerned,
we have assumed a \textit{smeared-out} one in contrast to real experimental situation where
it is \textit{discrete}. Previous numerical studies 
\cite{Messina_PhysicaA_2002,Elshad_2002,Moreira_EPL_2002}
have shown that the counterion distribution at inhomogeneously charged substrates
may deviate from that obtained at smeared-out ones at strong Coulomb coupling 
(i.e., multivalent counterions and/or high Bjerrum length) or
strong substrate charge modulations.
Nonetheless, at standard Bjerrum length (i.e., $l_B=7.1$  \AA\ for water at room temperature 
as is presently the case) and with discrete monovalent ions generating
the substrate's surface charge, it has been demonstrated that
the counterion distribution is marginally modified \cite{Messina_PhysicaA_2002}
even for trivalent counterions.
Hence, we think that our results will not qualitatively differ from the more realistic situation
of non-smeared-out substrate charges consisting of discrete monovalent ions.

An other approximation in our model is the location of the dielectric discontinuity.
More precisely, it was implicitly assumed that the latter coincides with 
the charged interface (considered here as a hard wall). 
In fact, experimentally, it is not clear where the dielectric discontinuity is located
and the transition is rather gradual and spreads out over several Angstr\"oms
\cite{Linse_JPC_1986},
so that in a continuum description the  dielectric discontinuity might be 
located somewhat below the \textit{hard} interface.
In this respect our model tends
to slightly overestimate the effect of image forces and namely, with $\Delta_{\epsilon}>0$,
the self-image \textit{repulsion}. 
Furthermore, in the presence of \textit{short-range} attractive interactions between the substrate and
the PEs (for instance stemming from some specific chemical properties of the chains and the substrate,
i.e. chemisorption),
the effect of image charges might also be reduced \cite{Dobrynin_JPCB_2003}.
This means that the substrate-charge \textit{undercompensation} by PEs induced by repulsive image forces 
as reported in Fig. \ref{fig.Qz_star_Qp-64}(b) is dependent on the relative strength of 
that short-range attractive interaction, which is not taken into account in our model.
Nevertheless, we are confident that our results provide a reliable fingerprint for the 
understanding of the effect of image forces on PE adsorption in a salt-free environment.

It is not a straightforward task to access experimentally to these effects stemming
from image forces. One major difficulty arises from the fact that by changing the 
dielectric constant of the solvent, $\epsilon_1$, one changes the degree of ionization of the PEs. 
However, there is the experimental possibility to tune $\Delta_{\epsilon}$ by using 
\textit{organic} solvents (i.e., with a low $\epsilon_1$ but still polar) 
with a mixture of large colloidal particles 
[e.g., latex particles with  weak curvature and  (low) dielectric constant $\epsilon_2$ such that 
$\epsilon_2 \leq \epsilon_1$] and PEs. In this experimental context, one should
be able to verify the trends of our current findings.


To conclude, we have performed MC simulations to address the effect of 
image forces on PE adsorption at oppositely charged planar substrates.
The influence of chain length and surface-charge density was also considered.
We have considered a finite monomer concentration in the dilute regime for relatively short chains.
Our main findings can be summarized as follows:
\begin{itemize}
\item 
For very short chains (here $N_m\leq 4$) and with no image forces (i.e., $\Delta_{\epsilon}=0$), 
the PE adsorption is similar to that occurring with little (spherical) multivalent counterions. \\
For longer chains (here $N_m\geq 8$), the PEs experience (even at $\Delta_{\epsilon}=0$) 
a short range repulsion near the substrate due to chain entropy effects. 
This latter feature is especially relevant at low substrate charge $\sigma_0$. 
\item 
At fixed $\sigma_0$ and in the presence of repulsive \textit{image forces} 
(here $\Delta_{\epsilon}=0.951$), it was demonstrated that the
monomer depletion in the vicinity of the substrate as well as 
the thickness of the PE layer grow with chain length $N_m$. 
Concomitantly, the \textit{charge reversal} of the substrate by the
adsorbed PEs \textit{vanishes}. 
\item 
Upon varying  $\sigma_0$ at fixed $N_m$, it was shown at  $\Delta_{\epsilon}=0$ that the 
net substrate-PE force becomes purely attractive at sufficiently high  $\sigma_0$, 
where chain-entropy effects are overcompensated.
When image forces are present, the PE \textit{depletion} near the substrate 
as well as the thickness of the adsorbed PE layer decrease with $\sigma_0$.
\end{itemize}

A future work will address the adsorption of stiff \textit{rod}-like PEs.
This very interesting situation was recently theoretically examined 
by Cheng and de la Cruz \cite{Cheng_JCP_2003}. 
Numerical simulation data would be of great help to further characterize the transversal and 
in-plane structures as well as to elucidate the influence of image forces.  

\acknowledgments 
The author thanks H. L\"owen for enlightening discussions.  
The SFB TR6 is acknowledged for financial support.


\begin{thebibliography}{10}
\bibitem{note_HS}
Only the monomer-monomer excluded volume interaction was not modelled by a
  hard-sphere potential. There, a purely repulsive Lennard-Jones potential was
  used.
\bibitem{note_overscreening}
Near the wall, the presence of the counterions at $\Delta_{\epsilon}=0$ is
  marginal so that the fluid charge is quasi exclusivly provided by the charges
  of the adsorbed monomers. This becomes especially vivid when $N_m \geq 4$
  where overcharging occurs.
\end{thebibliography}

\end{document}